\documentclass[preprintnumbers,prd,onecolumn,floatfix,superscriptaddress, nofootinbib]{revtex4-2}
\usepackage{setspace}
\usepackage{graphicx}
\usepackage{subcaption}
\usepackage{epsfig}
\usepackage{bm}
\usepackage{amssymb}
\usepackage{float}
\usepackage{amsmath}
\usepackage{dcolumn}
\usepackage[usenames,dvipsnames]{color}
\usepackage{enumitem}
\usepackage[mathscr]{eucal}
\usepackage{hyperref}
\def\doi{http://doi.org}

\newcommand{\be}{\begin{equation}}
\newcommand{\ee}{\end{equation}}
\newcommand{\beano}{\begin{eqnarray*}}
\newcommand{\eeano}{\end{eqnarray*}}
\newcommand{\ba}{\begin{eqnarray}}
\newcommand{\ea}{\end{eqnarray}}

\def\chariteratehelpA#1 #2\relax{%
  \chariteratehelpB#1\relax\relax%
  \ifx\relax#2\else\rlap{\charop{~}}\ \chariteratehelpA#2\relax\fi
}
\def\chariteratehelpB#1#2\relax{%
  \charop{#1}%
  \ifx\relax#2\else
    \chariteratehelpB#2\relax%
  \fi
}
\def\charop#1{\def\stacktype{L}\def\useanchorwidth{T}%
  \stackon[0pt]{#1}{\scalebox{.85}[1]{\color{red}$\sim$}}}

\usepackage{xcolor, soul}
\sethlcolor{yellow}

\begin{document}

\title{Late time phantom characteristic of the model in $f(R,T)$ gravity with quadratic curvature term}

\author{Shaily} 
\email{shaily.ma19@nsut.ac.in}
\affiliation{Department of Mathematics, Netaji Subhas University of Technology, New Delhi 110078, India}
\affiliation{School of Computer Science Engineering and Technology, Bennett University, Greater Noida 201310, India}
\author{Akanksha Singh} 
\email{akanksha.ma19@nsut.ac.in}
\affiliation{Department of Mathematics, Netaji Subhas University of Technology, New Delhi 110078, India}
\author{J. K. Singh}
\email{jksingh@nsut.ac.in}
\affiliation{Department of Mathematics, Netaji Subhas University of Technology, New Delhi 110078, India}
\author{Saibal Ray}
\email{saibal.ray@gla.ac.in}
\affiliation{Centre for Cosmology, Astrophysics and Space Science (CCASS), GLA University, Mathura 281406, Uttar Pradesh, India}

\begin{abstract}
\qquad We propose a novel cosmological framework within the $f(R,T)$ type modified gravity theory, incorporating a non-minimally coupled with the higher order of the Ricci scalar ($R$) as well as the trace of the energy-momentum tensor ($T$). Therefore, our well-motivated chosen $f(R,T)$ expression is $ R + R^m + 2 \lambda T^n$, where $\lambda$, $m$, and $n$ are arbitrary constants. Taking a constant jerk parameter ($j$), we derive expressions for the deceleration parameter ($q$) and the Hubble parameter ($H$) as functions of the redshift $z$. We constrained our model with the recent Observational Hubble Dataset (OHD), $Pantheon$, and $ Pantheon $ + OHD datasets by using the analysis of Markov Chain Monte Carlo (MCMC). Our model shows early deceleration followed by late-time acceleration, with the transition occurring in the redshift range $1.10 \leq z_{tr} \leq 1.15$. Our findings suggest that this higher order model of $f(R,T)$ gravity theory can efficiently provide a dark energy model for addressing the current scenario of cosmic acceleration.

\end{abstract}

\maketitle
PACS numbers: 04.20.-q, 04.50.Kd, 98.80.-k, 98.80.Es\\
Keywords: FLRW metric, $f(R,T)$ gravity, constant jerk parameter, phantom model

\section{Introduction}\label{intro}
Cosmological discoveries have revealed that astronomical structures in the universe exhibit a systematic arrangement rather than a random distribution. By studying this organized pattern and examining the physical characteristics of interstellar objects, we can gain insights into the phenomenon of cosmic accelerated expansion. While general relativity serves as the foundation of modern physics and effectively explains black holes and gravitational phenomena, it falls short of providing a comprehensive explanation for cosmic acceleration. To shed light on this intriguing concept, researchers have turned their attention towards investigating dark energy and exploring modified theories of gravity. Various astronomical studies, which include the Sloan Digital Sky Survey (SDSS), the 2-Degree Field Galactic Redshift Survey (2DFGRS), and the Huge Synoptic Survey Telescope (HSST), have demonstrated the significant influence that galaxies and stars have on the evolution of the universe. Analyzing these fundamental components is crucial to deepening our understanding of the formation and nature of the universe. Through the examination of gravitating systems, we can uncover the underlying physics of dark matter and dark energy. As theoretical cosmology confronts the challenges posed by the existence of dark energy and compelling experimental evidence, modified gravity theories emerge as encouraging avenues for refining our theoretical understanding of late-time cosmic acceleration.

Prominent modified gravitational theories in astrophysics as well as cosmology include $f(R)$~\cite{Capozziello:2002rd,Nojiri:2003ft,Carroll:2003wy,Bertolami:2007gv}, $f(R,T)$~\cite{Harko:2008qz,Harko:2014pqa, ShaHar1}, $f(T)$~\cite{Bengochea:2008gz,Linder:2010py,Boehmer:2011gw}, $f(G)$~\cite{Bamba:2010wfw,Houndjo:2013us}, $f(R,G)$~\cite{Nojiri:2005jg, ShaHar}, $f(G,T)$~\cite{Sharif:2016xjv,Shamir:2018qhq} and $f(R,\phi)$~\cite{Myrzakulov:2015qaa} (however, for interested readers, there are a few more references herein to consult the alternative gravity theories ~\cite{Li:2010cg,Capozziello:2011nr,Das:2017rhi,Deb:2017rhc,Deb:2018sgt,Deb:2018gzt,Deb:2017rhd,Biswas:2018inc,Biswas:2020gzd,Biswas:2021wfn,Tripathy:2021vjt,Maurya:2021aio,Mishra:2020jjk,Malik:2021dub,Singh:2022eun}). A cornerstone in the exploration of gravitational theories is the $f(R)$ gravity theory, pioneered by Buchdahl~\cite{Buchdahl:1970ldb}, which involves substituting the Ricci scalar $R$ in the Einstein-Hilbert (EH) action with an arbitrary function $f(R)$. This theory has been examined and verified as a viable explanation for late-time cosmic acceleration while also aligning with local gravitational tests~\cite{Carroll:2003wy,Nojiri:2007as,Nojiri:2007cq,Cognola:2007zu,Elizalde:2010ts,Nojiri:2010wj,Chakraborty:2014xla}. While $f(R)$ gravity remains a focus of extensive study, recent decades have witnessed a surge of interest in $f(T)$ gravity. This alternative theory has also provided explanations regarding cosmic expansion and galactic dynamics, as evidenced by references~\cite{Rodrigues:2012qua,Salako:2013gka,Paliathanasis:2016vsw}. In their work, Harko et al.~\cite{Harko:2011kv} first put forward the $f(R,T)$ theory as a modification of general relativity (GR) that involves altering the geometric part of the EH action, where $R$ is the Ricci scalar and $T$ denotes the trace of the energy-momentum tensor (EMT). In this theory, $R$ is replaced by an arbitrary $f(R,T)$ function, and due to the incorporation of $T$, this theory encompasses novel gravitational aspects and also satisfies the weak-field conditions of the solar system.

In the realm of cosmic dynamics, Houndjo~\cite{Houndjo:2011tu} shed light on the transition from a matter-dominated epoch to the late-time acceleration phase using the minimally coupled $f(R,T)$ model. Within the vast expanse of scientific literature, several distinct formulations emerge, each offering a unique lens into the functional dependence of $f(R,T)$. Researchers have explored various mathematical expressions to represent the dependence on $T$, including polynomials, rational functions, and exponential expressions~\cite{Yousaf:2016lls,Shabani:2014xvi,Shabani:2013djy,Moraes:2015uxq,Roshan:2016mbt,Noureen:2015nja,Moraes:2019hgx}. It is argued that the $T$-dependence can arise due to the presence of quantum effects or imperfect fluids. Amidst the array of modified gravitational theories, the $f(R,T)$ theory stands out as it incorporates an energy transfer relationship between matter and geometry, thus potentially exhibiting the observed phenomenon related to the accelerating phase of the cosmic expansion. In connection to a few important features of $f(R,T)$ gravity theory, the following works are notable~\cite{Shamir:2015rva,Singh:2014bha,Jamil:2011ptc,Correa:2015qma,Sharif:2012zzd}. The consequences of the $f(R,T)$ theory on cosmological and solar system models have been investigated in~\cite{Deng:2014hga}. In literature, discussions on bouncing models within the $f(R,T)$ theory have also gained attention~\cite{Singh:2023gxd,Singh:2022jue,Singh:2018xjv}.

Motivated by the above-mentioned $f(R,T)$ theory-based works, our current investigation starts with the general powers of curvature and trace of the energy-momentum tensor. Specifically, we consider the form $f(R, T) = R + R^m + 2 \lambda T^n$, subsequently delving into a detailed analysis with distinct parameter values for $m$ and $n$. In this context, it is to be noted that for wormhole-related investigations, Elizalde and Khurshudyan~~\cite{Elizalde:2018arz} exploited a similar form of $f(R,T)$ with specific parametric values, such as $m=2$ and $n=1$. We also notice that very recently Mahapatra and Das~\cite{Mahapatra:2024kfq} have employed a polytropic model under $f(R,T)$ in a restricted manner to study the equilibrium configurations of relativistic neutron stars. However, in this connection, we would like to emphasize that our approach is more general so that the results can be expected to offer a wide range of flexibility to explore the multifarious features of the late-time phantom behavior of the universe.

The structured sequence of our research is: Section \ref{FormulationHz} serves as the starting point, where we delve into the computation of the Hubble parameter by using the jerk and deceleration parameter expressions. Next, we derive the Einstein field equations for a specific form of the $f(R,T)$ function in Sec. \ref{sectionEFE}. In our model, we constrain three model parameters using the OHD dataset, the $ Pantheon $ dataset, and their joint (OHD + $ Pantheon $) dataset in Sec. \ref{observational}. Sec. \ref{dynamicalparameters} contains the explanation related to the dynamical behavior of the universe for this model. At last, in Sec. \ref{results}, we discuss the results and conclude our work.

\section{Formulation of the Hubble parameter}\label{FormulationHz}

\qquad Let us first define here the jerk parameter $j$
\begin{equation}\label{1}
j(z) = q(z)+2q(z)^2+(1+z)\frac{dq(z)}{dz},
\end{equation}
where $q$ and $z$ are, respectively, the deceleration parameter and the redshift.

Treating $j$ as a constant and subsequently solving the above differential equation leads to
\begin{equation}\label{2}
\frac{\sqrt{8j+1}+4q+1}{\sqrt{8j+1}-4q-1} = \left(\alpha+\alpha z\right)^{\sqrt{8j+1}},
\end{equation}
where $ \alpha $ is an integration constant.

Rewriting Eq. (\ref{2}) we get
\begin{equation}\label{3}
q = \frac{-1}{4} + \frac{\sqrt{8j+1}}{4} - \frac{\sqrt{8j+1}}{2\left(1+\left(\alpha+\alpha z\right)^{\sqrt{8j+1}}\right)}.
\end{equation}

The equation representing $q$ as a function of the Hubble parameter $H$ and redshift $z$ is
\begin{equation}\label{4}
q(z) = -1 +\frac{(1+z)}{H(z)}\frac{dH}{dz}.
\end{equation}

Using Eq. (\ref{3}) in Eq. (\ref{4}), we get
\begin{equation}\label{5}
\frac{1}{1+z}\left[ 1 - \frac{1}{4} + \frac{\sqrt{8j+1}}{4} - \frac{\sqrt{8j+1}}{2\left(1+\left(\alpha+\alpha z\right)^{\sqrt{8j+1}}\right)} \right] dz = \frac{dH}{H}.
\end{equation}

By integrating both sides of Eq. (\ref{5}), we obtain
\begin{equation}\label{6}
H = \beta \left(1+z\right)^{\frac{3+\sqrt{8j+1}}{4}} \sqrt{1+\left(\alpha+\alpha z\right)^{-\sqrt{8j+1}}},
\end{equation}
where $ \beta $ is an integration constant.

In order to work with observational data, we define $ \beta $ as $ \frac{H_0}{\sqrt{1+\alpha^{-\sqrt{8j+1}}}} $, where $ H_0 = H(0) $. Subsequently, one can find the $ H $, which is as follows
\begin{equation}\label{7}
H = \frac{H_0}{\sqrt{1+\alpha^{-\sqrt{8j+1}}}} \left(1+z\right)^{\frac{3+\sqrt{8j+1}}{4}} \sqrt{1+\left(\alpha+\alpha z\right)^{-\sqrt{8j+1}}}.
\end{equation}

\section{Formulation of $f(R,T)$ theory of gravity}\label{sectionEFE}

\qquad The gravitational action corresponding to $f(R,T)$ gravity theory is
\begin{equation}\label{8}
\mathscr{S}=\int \Bigg[\frac{f(R,T)}{16 \pi G}+\mathscr{S}_m \Bigg]\sqrt{-g}  d^{4}x,
\end{equation}
where $R$, $T$, $g$, $G$, and $\mathscr{S}_m$, respectively, denote the Ricci scalar, the trace of the EMT, the determinant of the metric tensor, the gravitational constant, and the matter Lagrangian. Here, the function $f(R,T)$ is explicitly defined as $f(R,T)=f_1(R)+2f_2(T)$~\cite{Harko:2011kv}.

On varying the action (\ref{8}) with respect to the metric tensor $g_{\mu\nu}$, one can easily get the following gravitational equation~\cite{Singh:2022eun}
\begin{equation}\label{9}
f_1^R (R) R_{\mu\nu} - \frac{1}{2} g_{\mu\nu}(f_1(R)+2f_2(T))+(g_{\mu\nu} \Box -\nabla_\mu \nabla_\nu) f_1^R(R)=8\pi G T_{\mu\nu}-2f_2^T(T)(T_{\mu\nu}+\Theta_{\mu\nu}),
\end{equation}
where $ f_1^R (R) $ symbolizes the first order differentiation of $ f_1 $ with respect to $ R $ and $ f_2^T(T) $ means the first order differentiation of $ f_2 $ with respect to $ T $. The d'Alembert operator is denoted by $\Box$, and $\nabla_\mu$ signifies the covariant derivative with respect to $g_{\mu\nu}$. The expression $\Theta_{\mu\nu}$ can be written as
\begin{equation}\label{10}
\Theta_{\mu\nu}=g^{ij} \frac{\delta{T_{ij}}}{\delta{g_{\mu\nu}}}=-2T_{\mu\nu}+g_{\mu\nu} \mathscr{S}_m-2g^{ij} \frac{\delta^2 \mathscr{S}_m}{\delta{g_{\mu\nu}} \delta{g_{ij}}}.
\end{equation}

Considering a perfect fluid, the matter Lagrangian is conventionally represented as $ \mathscr{S}_m =-p $. In this case, the EMT is given by $ T_{\mu\nu} = (\rho+p)u_\mu u_\nu-p g_{\mu\nu} $, where the physical parameters have their usual meanings. Using these expressions, we get $ \Theta_{\mu\nu}=-2 T_{\mu\nu}-p g_{\mu\nu} $. Therefore, Eq. (\ref{9}) takes the following form
\begin{equation}\label{11}
f_1^R (R) R_{\mu\nu} - \frac{1}{2} g_{\mu\nu}(f_1(R)+2f_2(T))+(g_{\mu\nu} \Box -\nabla_\mu \nabla_\nu) f_1^R(R)=8\pi G T_{\mu\nu}+2f_2^T(T)(T_{\mu\nu}+p g_{\mu\nu}).
\end{equation}

The form in which the Einstein field equations are expressed is given by
\begin{equation}\label{12}
G_{\mu\nu}=R_{\mu\nu}-\frac{1}{2} R g_{\mu\nu}= \frac{8 \pi G}{f_1^R(R)}(T_{\mu\nu}+T^*_{\mu\nu}),
\end{equation}
where \\
$ ~~~~~~~~~~T^*_{\mu\nu}=\frac{1}{8 \pi G}\Big[\frac{1}{2} g_{\mu\nu}((f_1(R)+2f_2(T))-R f_1^R (R))+(\nabla_\mu \nabla_\nu-g_{\mu\nu} \Box) f_1^R(R)+2f_2^T(T)(T_{\mu\nu}+p g_{\mu\nu})\Big].$ \\
   
Now, the metric of the FLRW universe is expressed by 
\begin{equation}\label{13}
ds^2=dt^2-a^2 (t)\sum\limits_{i=1}^{3}dx_i^2.
\end{equation}

Here, $ a(t) $ is the scale factor. Now, by considering $ f_1(R)=R+R^m $ and $ f_2(T)=\lambda T^n $, where $ T=\rho-3p $ and $ R=-6(2 H^2+\dot{H}) $, we can obtain the Einstein field equations (EFE) as
\begin{equation}\label{14}
3 H^2 = \frac{1}{1+ mR^{m-1}} [ 8\pi\rho + \frac{1}{2} ((1-m)R^m+2\lambda T^n) - 3H m(m-1) R^{m-2} \dot{R} + (\rho +p)2\lambda n T^{n-1} ]	\nonumber
\end{equation}
or
\begin{eqnarray}\label{15}
3 H^2 = \frac{1}{1+ mR^{m-1}} [ 8\pi\rho + \frac{1}{2} ((1-m)(-6(2H^2+\dot{H}))^m+2\lambda T^n) ] + ~~~~~~~~~~~~~~~~~~~~~~~~~~   \nonumber   \\
\frac{1}{1+ mR^{m-1}} [ - 3H m(m-1) (-6(2H^2+\dot{H}))^{m-2} (-6(4H\dot{H}+\ddot{H})) + (\rho +p)2\lambda n T^{n-1} ],	\nonumber
\end{eqnarray}

\begin{equation}\label{16}
2\dot{H} + 3H^2 = \frac{-1}{a^2 (1+ mR^{m-1})} [8\pi pa^2 + \frac{1}{2} (-a^2) ((1-m)R^m+2\lambda T^n ) + a m(m-1)R^{m-3} (a ((m-2)\dot{R}^2 + R\ddot{R}) + 3\dot{a}R\dot{R} ) ]	\nonumber
\end{equation}
or
\begin{eqnarray}\label{17}
2\dot{H} + 3H^2 = \frac{-1}{a^2 (1+ mR^{m-1})} [8\pi pa^2 + \frac{1}{2} (-a^2) ((1-m)(-6(2H^2+\dot{H}))^m+2\lambda T^n )+ ~~~~~~~~~~~~~~~~~~~~~~~~~~~~~~~~~~~~~~~~~~~~~~~~~~~~~~~~~~~~~~~~~~~~~~~~~ \notag \\   a\zeta m(m-1)(-6(2H^2+\dot{H}))^{m-3} (a ((m-2)(-6(4H\dot{H}+\ddot{H}))^2 + (-6(2H^2+\dot{H}))(-6(4H\ddot{H}+4\dot{H}^2+\dddot{H}))) + ~~~~~~~~~~~~~~~~~~~~~~~~~~~~~~~~~~~~\notag \\ 3\dot{a}(-6(2H^2+\dot{H}))(-6(4H\dot{H}+\ddot{H})) ) ].~~~~~~~~~~~~~~~~~~~~~~~~~~~~~~~~~~~~~~`	\nonumber
\end{eqnarray}

As the calculations involved in determining the physical parameters $ \rho $ and $ p $ based on the above equations are complex, we simplify the analysis by choosing $ m = 2 $ and $ n = 1 $. Now, we have taken $ f_1(R)=R+R^2 $, a choice previously addressed by Starobinsky~\cite{Starobinsky:1980te}, and $ f_2(T)=\lambda T $. The EFEs for our chosen model are
\begin{equation}\label{18}
(8 \pi+3\lambda)\rho-\lambda p=3 H^2 +18 (\dot{H}^2-6 H^2 \dot{H}-2 H \ddot{H}),
\end{equation}
\begin{equation}\label{19}
(8 \pi+3 \lambda ) p-\lambda \rho=-2\dot{H}-3 H^2+6 (26 \dot{H} H^2+9\dot{H}^2+14 H \ddot{H}+2\dddot{H}).
\end{equation}

\section{Observational analysis} \label{observational}

\qquad Observational analysis is an endeavoring field in cosmology where we calculate the optimal value of model parameters with the help of some experiments. In this analysis, the most accurate and strong persistence results are obtained from the Planck satellite mission. In this model, we determine the optimal values for model parameters by conducting a best-fit analysis with the help of three distinct observational datasets, namely, the Hubble dataset (OHD) with $ 77 $ data points, the $ Pantheon $ dataset with $ 1048 $ data points, and the combined dataset (OHD + $ Pantheon $). The Hubble parameter value from Eq. (\ref{7}), which contains three model parameters named $\alpha$, $j$, and $H_0$, is used for our analysis. In order to constrain the mentioned model parameters, we basically execute the Markov Chain Monte Carlo (MCMC) method, implemented through the $emcee$ library in Python.

\subsection{$ H(z) $ Dataset}

\qquad Exploring the dark division of the universe involves tapping into the insightful $H(z)$ data, and this data is basically acquired from the cosmic chronometers~\cite{Chimento:2007da}. In our model, we use the OHD along with 77 data points~\cite{Shaily:2022enj}. Thereafter, by using the $ \chi^2 $ test, we find the parametric best-fit values of the present model.

Now, the formula for $ \chi^2 $ is drafted as	
\begin{equation}\label{20}
\chi _{Hb}^{2}(\alpha,j,H_0)=\sum\limits_{i=1}^{77} \frac{[H(\alpha,j,H_0,z_{i})-H_{obs}(z_{i})]^2}{\sigma _{z_i}^2},
\end{equation}
where $H(\alpha, ~j, ~H_0, ~z_{i})$ and $H_{obs}$ are denoted for, respectively, the theoretical and observed values of the Hubble parameter. Additionally, $ \sigma_{(z_{i})} $ stands for the standard deviation for every value of $H(z_i)$.

We determine that the analysis of only the Hubble dataset and the function $\chi_{Hb}^{2}$ is not very authentic for the cosmological models. Therefore, we take the $ Pantheon $ dataset and their joint dataset into consideration.

\begin{figure}\centering 
	\includegraphics[scale=0.54]{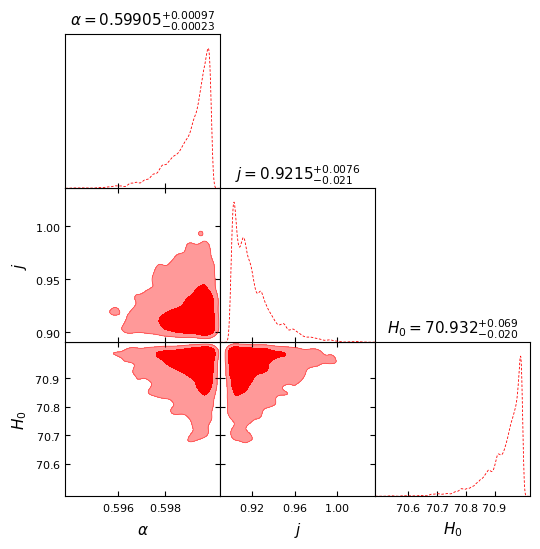} 	
\caption{Visualization of the likelihood contours using the $ H(z) $ dataset.}
\label{hzc}
\end{figure}

\begin{table}[H]
\caption{Best-fit values of model parameters where $D_1 = H(z)$, $D_2= Pantheon$, and $D_3= H(z)+Pantheon$.}
\begin{center}
\label{tabparm}
\begin{tabular}{l c c c r} 
\hline\hline
\\ 
{Dataset} &      ~~~~~  $ \alpha $  & ~~~~~  $ j $ & ~~~~~ $ H_0 $  
\\
\\
\hline      
\\
{$D_1$}     &  ~~~~~ $ 0.59905^{+0.00097}_{-0.00023} $  &  ~~~~~ $ 0.9215^{+0.0076}_{-0.0210} $   &  ~~~~~ $ 70.932^{+0.069}_{-0.020} $ 
\\
\\
{$D_2$}     &  ~~~~~ $ 0.59998 \pm 0.0001 $  &  ~~~~~ $ 0.949998 \pm 0.000098 $   &  ~~~~~ $ 70.000013^{+0.000081}_{-0.000099} $ 
\\
\\
{$D_3$}  &  ~~~~~ $ 0.598987^{+0.000094}_{-0.000110} $  &  ~~~~~ $ 0.98999^{+0.000120}_{-0.000085} $   &  ~~~~~ $ 70.3719 \pm 0.0001 $
\\
\\ 
\hline\hline  
\end{tabular}  
\end{center}
\end{table}

\subsection{Pantheon Dataset}
\qquad In the last two decades, Type Ia supernovae have emerged as pivotal tools in the analysis of cosmological parameters, notably serving as the first evidence for the late time accelerated expansion phase of the universe. Supernovae are the stars, and as they burst, they release a lot of energy and enlarge their outer shell. Thus, this part is related to examining the time evolution of their brightness and their spectrum. In this model, the $ Pantheon $ dataset which comprises of $1048$ data points, is used to constrain the parameters of the model~\cite{Pan-STARRS1:2017jku}. The $ Pantheon $ dataset contains the resulting compilation of different surveys, including the CfA1-CfA4 surveys, the Sloan Digital Sky Survey (SDSS), the Supernovae Legacy Survey (SNLS), the Carnegie Supernova Project (CSP), the Pan-STARRS1 (PS1), and various Hubble Space Telescope (HST) surveys, all of which span a redshift range from $0.01$ to $2.26$~\cite{Riess:1998dv,Jha:2005jg,Hicken:2009df,Contreras:2009nt,SDSS:2014irn}. In Type Ia supernovae analysis, we deal with redshift as well as luminosity distance, which are closely related to standard candles in the universe. The luminosity distance is defined as
\begin{equation}\label{21}
D_L(z)=(1+z) \int_0^z \frac{c}{H(z^*)}dz^*,
\end{equation}
with $c$ as the speed of light in a vacuum.

Now, $m$, the apparent magnitude can be provided as
\begin{equation}\label{22}
m(z)=M+5log_{10}\Bigg[\frac{D_L(z)}{1Mpc}\Bigg]+25,
\end{equation}
with $M$ denoting the absolute magnitude.

Here, a dissipation between $H_0$ and $M$ is observed, and given that the distance modulus $\mu$ is equal to $m-M$, the $\chi^2$ formula is written as
\begin{equation}\label{23}
\chi_{Pn}^{2}(\alpha,j, H_0)=\sum\limits_{i=1}^{1048}\left[ \frac{\mu_{th}(\alpha,j, H_0,z_{i})-\mu_{obs}(z_{i})}{\sigma _{\mu(z_{i})}}\right] ^2,
\end{equation}
where $\mu_{th}$ and $\mu_{obs}$ display the theoretical and observed distance moduli, respectively.

\begin{figure}\centering 
	\includegraphics[scale=0.54]{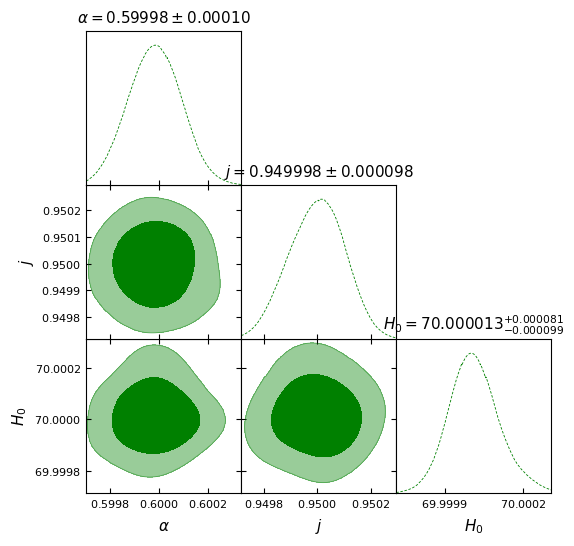} 	
\caption{Visualization of the likelihood contours for the $ Pantheon $ dataset.}
\label{panc}
\end{figure}

\subsection{Joint Dataset ($ H(z)+Pantheon $)}

\qquad For enhanced precision in constraining model parameters, we make use of the joint dataset, and the results are calculated by the $ \chi_{HP}^2 $ function, which is the sum of the functions $ \chi_{Hb}^{2} $ and $ \chi_{Pn}^{2} $ as
\begin{equation}\label{24}
\chi _{HP}^{2}(\alpha,j, H_0)=\chi _{Hb}^{2}(\alpha,j, H_0)+\chi _{Pn}^{2}(\alpha,j, H_0).
\end{equation}

\begin{figure}\centering 
	\includegraphics[scale=0.54]{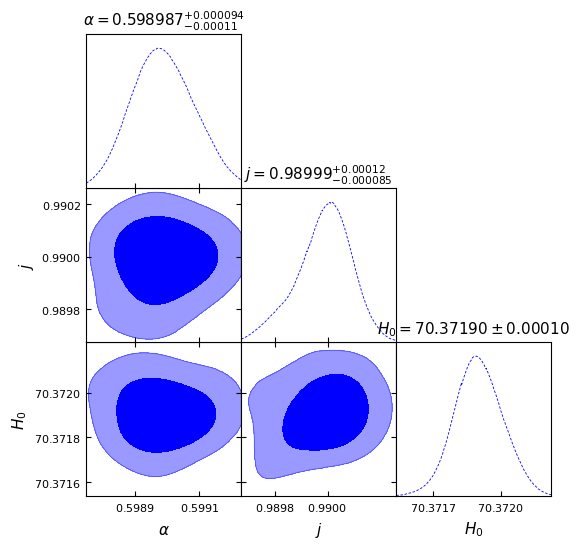} 	
\caption{Visualization of the likelihood contours for the $ H(z)+Pantheon $ dataset.}
\label{hzpanc}
\end{figure}

The obtained values of $\alpha$, $j$, and $H_0$ from the above-mentioned observational datasets are placed in Table \ref{tabparm}. In Fig. \ref{hzc}, we notice that the obtained contours are not properly in an oval shape, but we have tested using the Gelman-Rubin convergence test that such contour patterns also converge~\cite{Singh:2022nfm,Gelman:1992zz}.

\begin{figure}\centering
	\subfloat[]{\label{a}\includegraphics[scale=0.48]{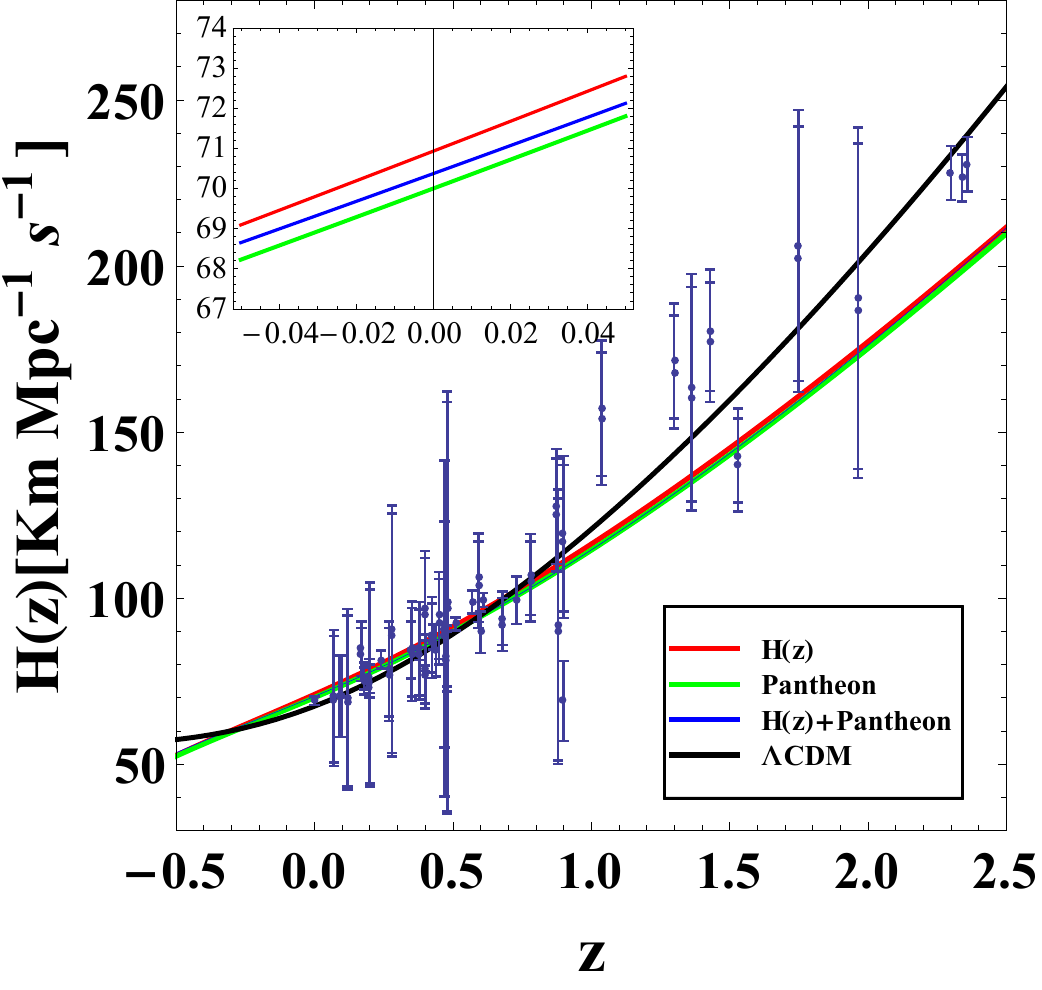}}\hfill
	\subfloat[]{\label{b}\includegraphics[scale=0.45]{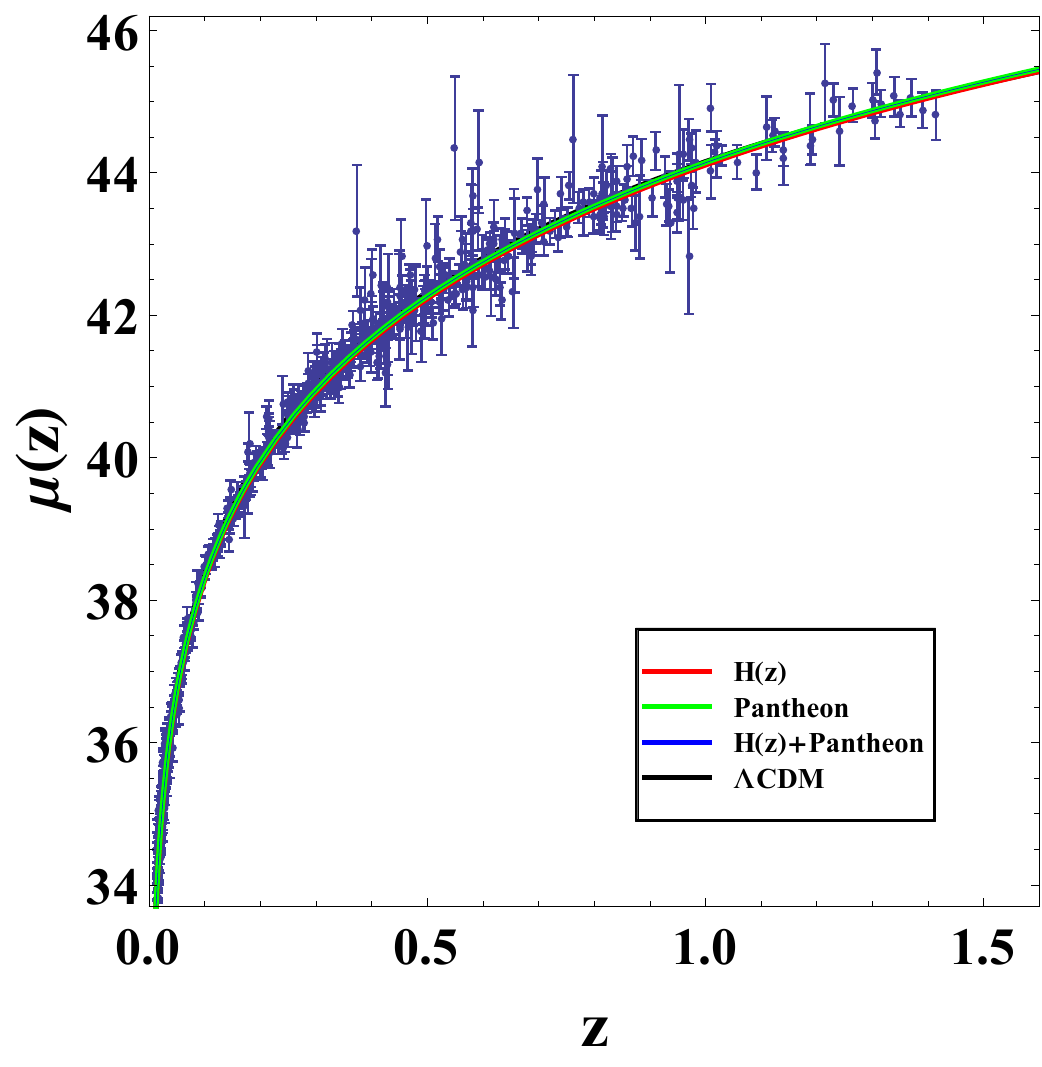}} 
\caption{The error bar plots for $ H(z) $ and Type Ia Supernovae datasets exhibiting the similarities between the presented model and the $ \Lambda $CDM.}
\label{err}
\end{figure}

For a better understanding of our model, we conduct a comparative analysis with the $ \Lambda $CDM model, illustrated through error bar plots. In Fig. \ref{err}, there are three distinct trajectories concerning the best-fit parametric values of the model. The Hubble dataset and the Type Ia Supernovae dataset based error plots are presented in Figs. \ref{err}(a) and \ref{err}(b), respectively.

\section{Physical parameters}\label{dynamicalparameters}
 
\qquad This section involves the scrutinization of physical parameters, systematically exploring their behavior across the best-fit values corresponding to the Hubble dataset, the $Pantheon$ dataset, and their joint dataset. To gain knowledge about phase-wise cosmic evolution, we examine the features of the different model parameters, especially the dimensionless deceleration parameter expressed in Eq. (\ref{3}).

\begin{figure}\centering
	\includegraphics[scale=0.46]{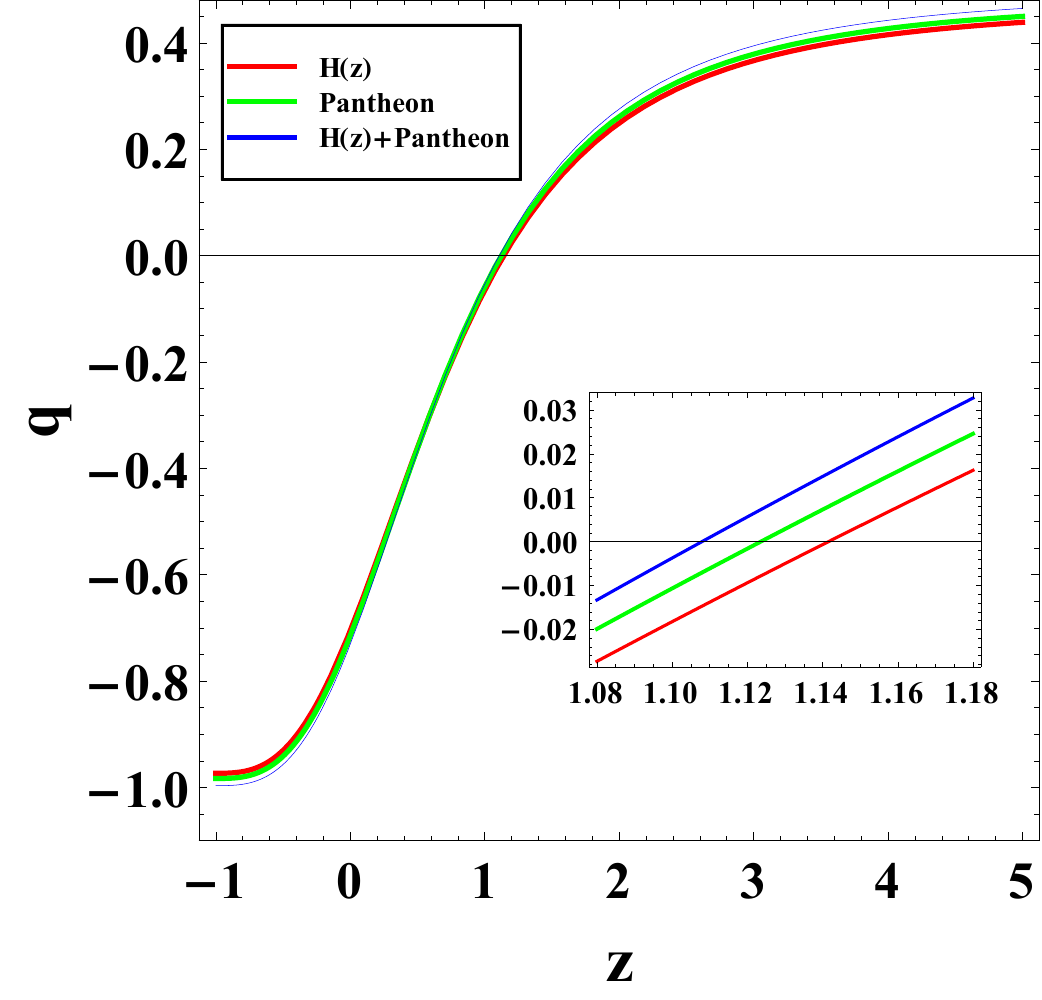}\hfill
\caption{ The graphical representation of $ q $ \textit{vs.} $ z $.}
\label{decp}
\end{figure}

Fig. \ref{decp} demonstrates a trend in which the deceleration parameter $q$ experiences a declination from high to low redshift values. Also, this parameter undergoes a sign change, indicating a phase shift from the deceleration to the cosmic acceleration. It is observed that for all the trajectories at the early epochs, the universe was in the deceleration mode, whereas at present, as well as in late time eras, the universe is in the accelerating mode. The values of redshift corresponding to the phase transition phenomena are $ 1.14161 $, $ 1.12361 $, and $ 1.10797 $ for the $ H(z) $ dataset, the $ Pantheon $ dataset, and their joint dataset, respectively. Furthermore, the deceleration parameter assumes values of $-0.705673$, $-0.715246$, and $-0.731007$ at the present for $ H(z) $, $Pantheon$, and their joint dataset, respectively.

For convenience,the jerk parameter $j$ can be articulated in relation to the deceleration parameter $q$, as in Eq. (\ref{1}). In the context of the $ \Lambda $CDM model, the jerk parameter maintains a constant value of $ 1 $. In our model, we determine the best-fit values for the jerk parameter to be $ 0.9215 $, $ 0.949998 $, and $ 0.98999 $ for the $ H(z) $ dataset, the $ Pantheon $ dataset, and their joint dataset, respectively. Notably, all three values of $ j $ are remarkably close to unity, reflecting the consistency and agreement found within the results.

\begin{figure}\centering 
	\subfloat[]{\label{a}\includegraphics[scale=0.48]{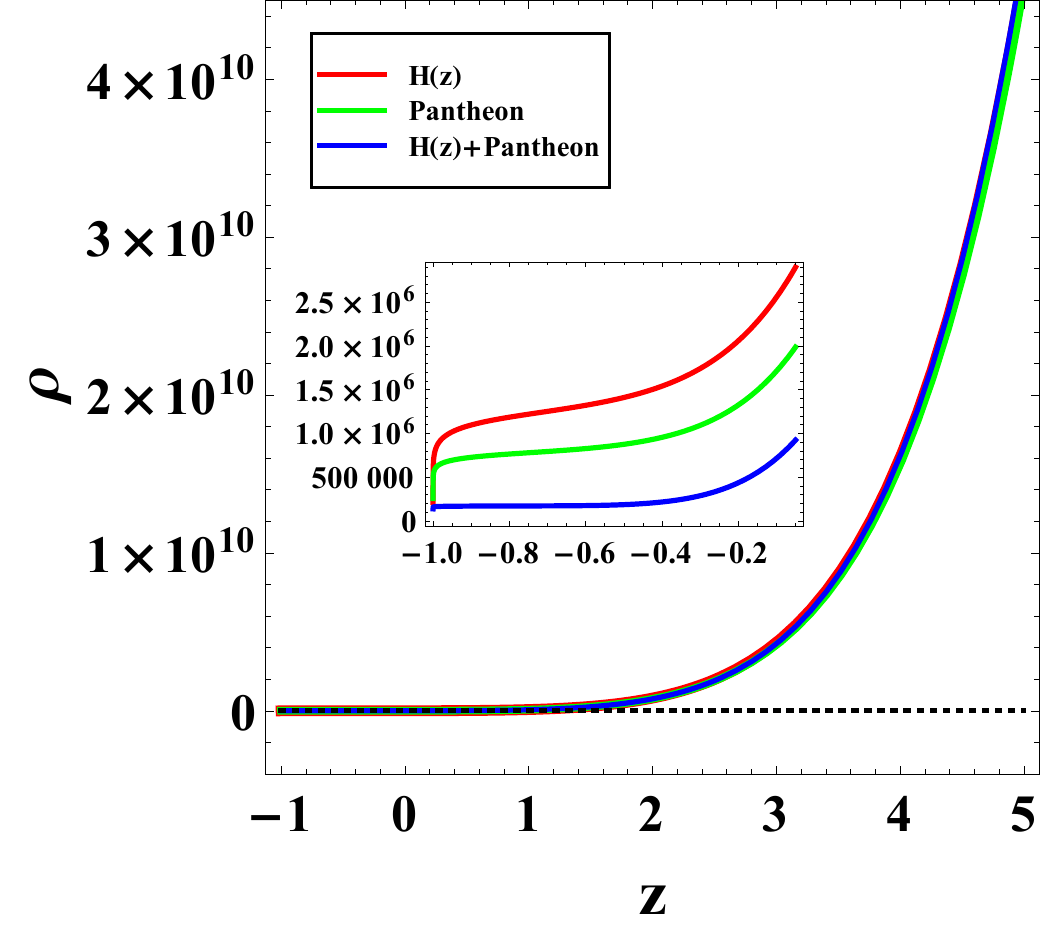}}	\hfill
	\subfloat[]{\label{b}\includegraphics[scale=0.47]{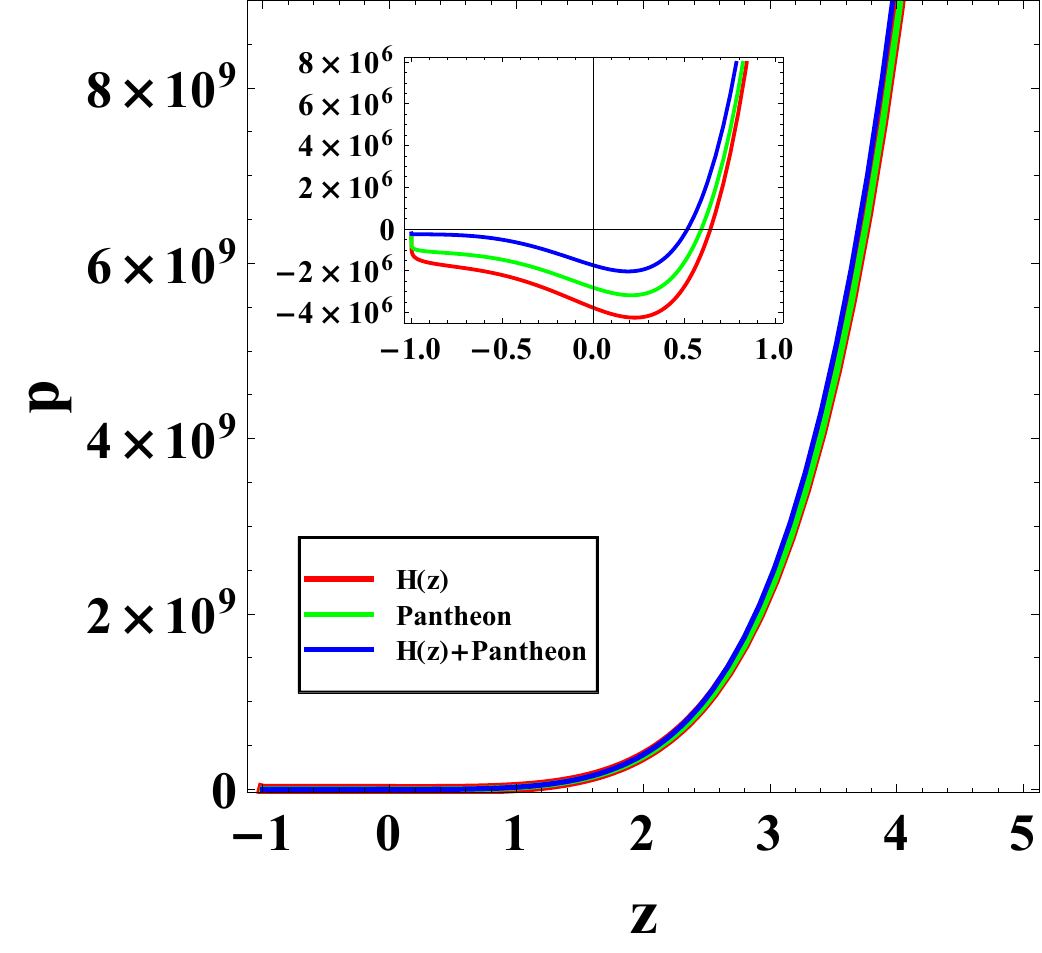}}\par
	\subfloat[]{\includegraphics[scale=0.45]{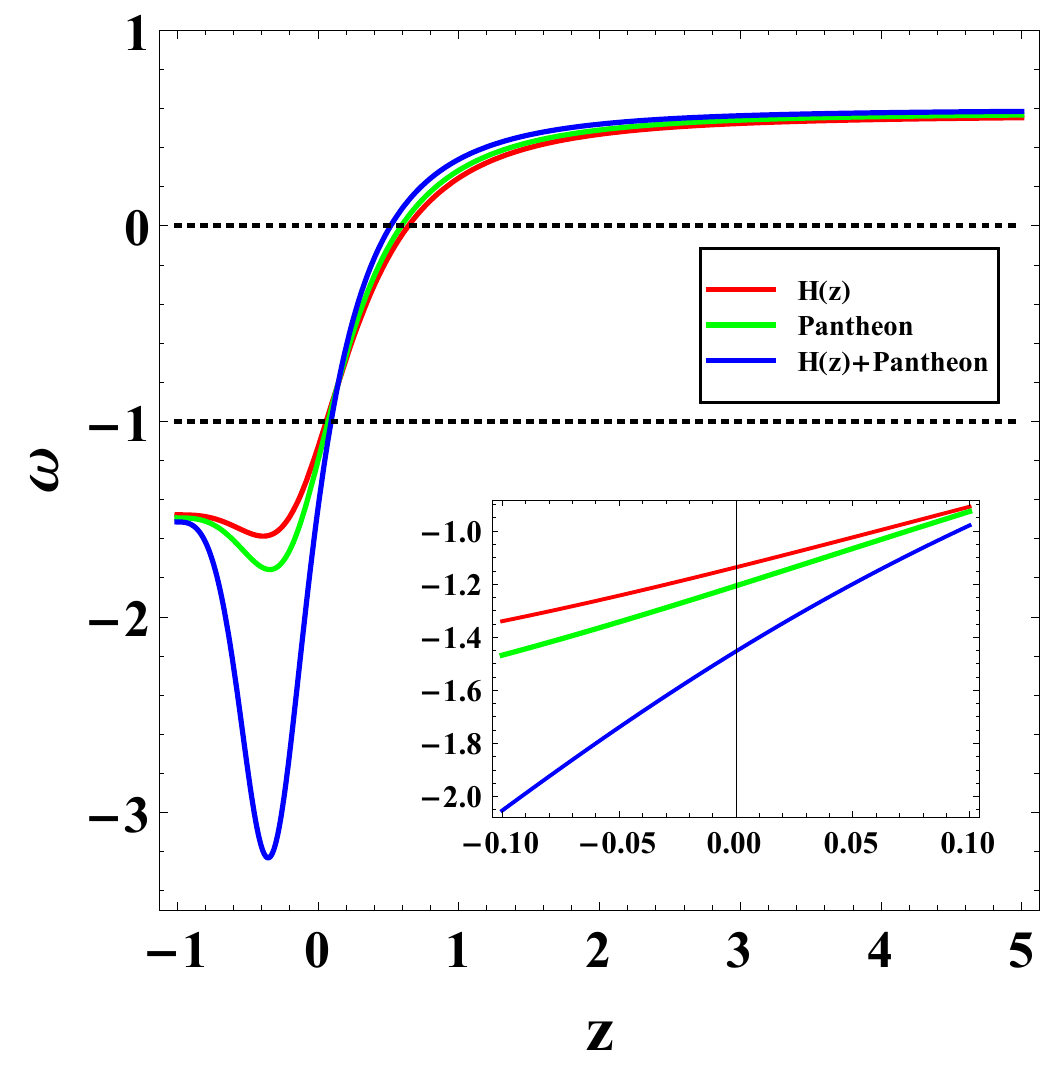}}
\caption{The graphical representations of the physical parameters, viz., $ \rho $, $ p $, and $ \omega $ \textit{vs.} $ z $ for all the observational datasets.}
\label{phypara}
\end{figure}

By using Eqs. (\ref{18}) and (\ref{19}), we calculate the values of $\rho$ and $p$. However, since their expressions are very large and complicated, we have plotted their trajectories directly for every mentioned dataset. We have Fig. \ref{phypara}(a), in which it is observed that $ \rho $ is decreasing monotonically from the early to late times. For the isotropic pressure, Fig. \ref{phypara}(b) shows that all the curves are in a highly positive region, but afterward, they decrease and enter a negative region. At present, the negative value of $p$ signifies an expanding and accelerating universe, while in late times, the model becomes pressureless since $p$ tends to zero as $z$ approaches $ -1 $. In Fig. \ref{phypara}(c), the progression of the Equation of State (EoS) parameter ($\omega = p/\rho$) has been unfolded: starting from a perfect fluid, then it is entering the quintessence region and eventually remaining in the phantom region. For all observational datasets, the value of $\omega$ is less than $ -1 $, indicating that our model represents a phantom scenario both at the present as well as in the late times.

Besides phenomenal fields, there are some major points to explore, such as the classification of singularities, causal structure, energy conditions, etc. In particular, energy conditions originated from the well-known Raychaudhuri equation~\cite{Raychaudhuri:1953yv,Ray1957,Bhattacharyya:2021djv,Das:2023umo,Panda:2023jwe}, serving as a valuable framework to assess the influence of all the geometric terms within the stress-energy tensor. Stemming from the Einstein field equations, energy conditions involve diminutions of time-like or null vector fields with respect to the EMT and Einstein tensor. These conditions stand out as prominent features in cosmological models, defined as the weak energy condition (WEC: $ \rho \geq 0,~ \rho+p > 0 $), null energy condition (NEC: $ \rho+p \geq 0 $), strong energy condition (SEC: $ \rho+3 p \geq 0 $), and dominant energy condition (DEC: $ \rho \geq |p| $).

\begin{figure}\centering 
	\subfloat[]{\label{a}\includegraphics[scale=0.45]{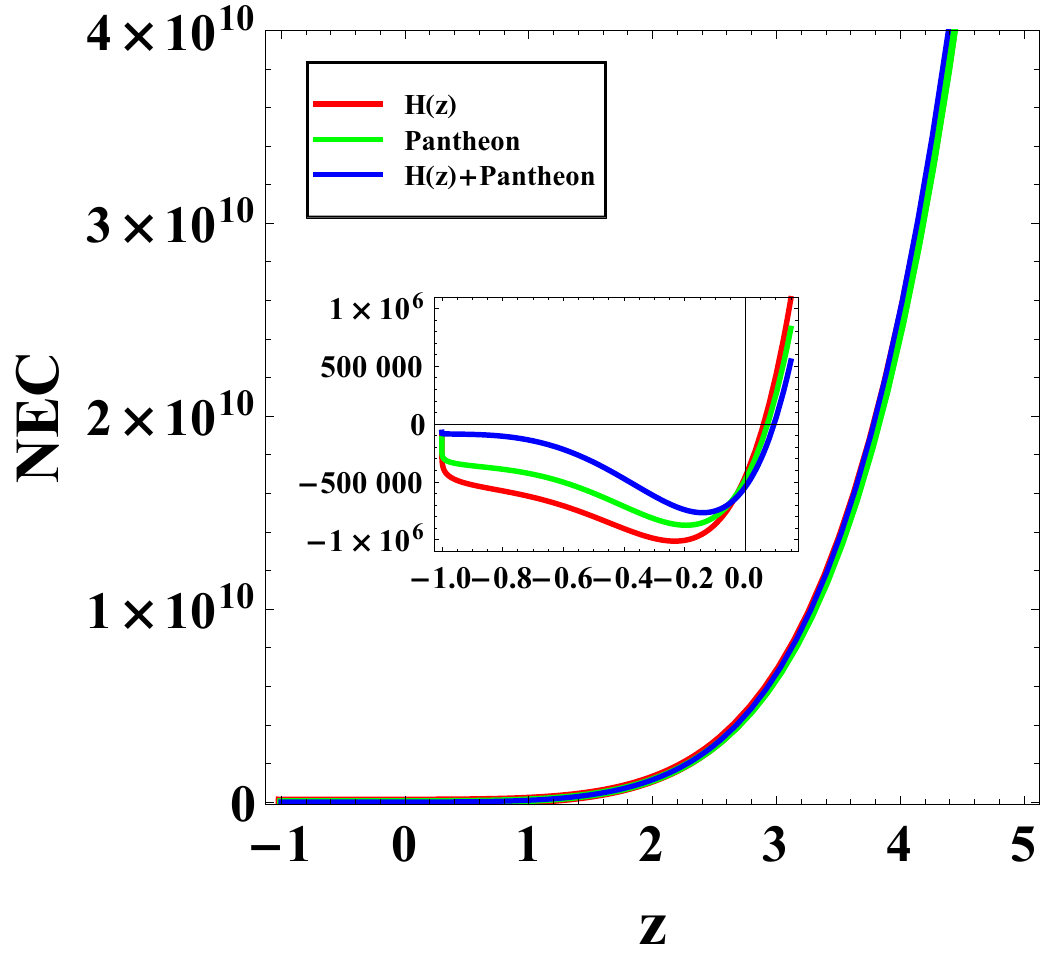}}\hfill
	\subfloat[]{\label{b}\includegraphics[scale=0.45]{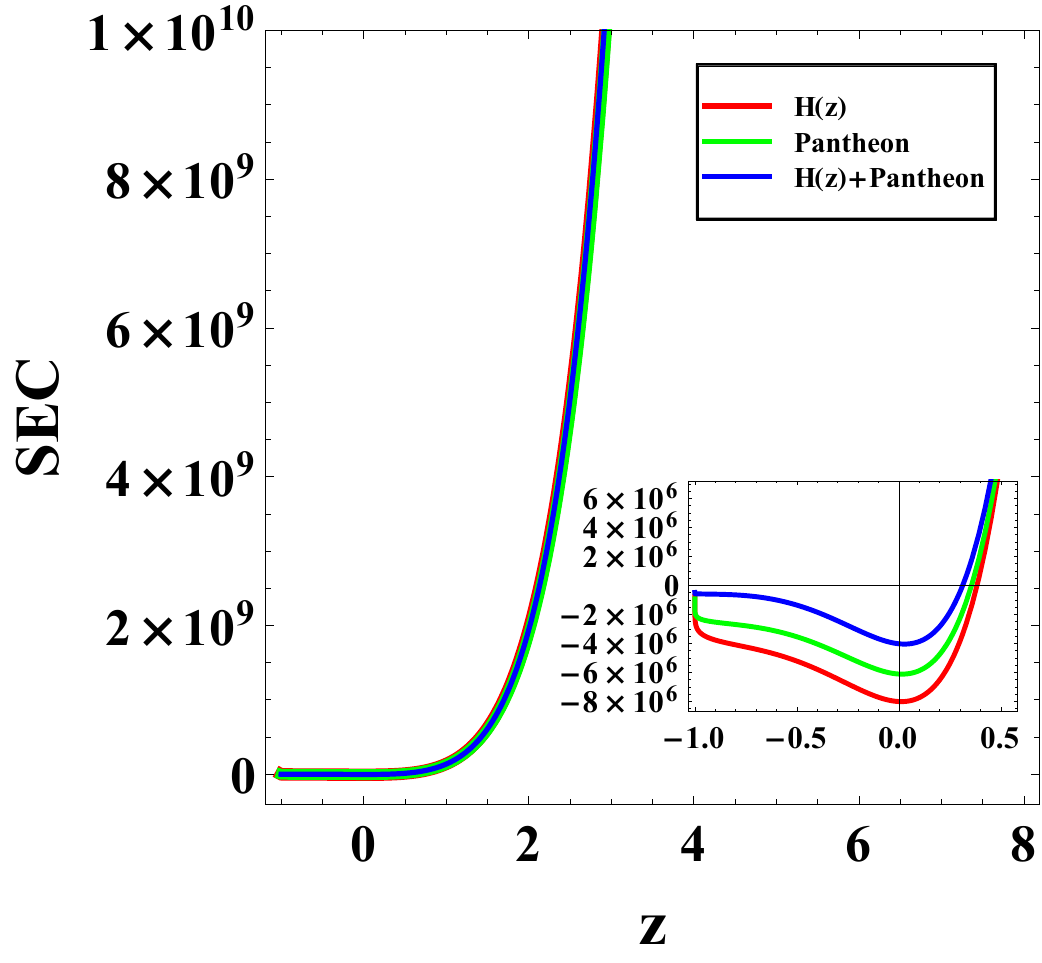}}\par 
	\subfloat[]{\label{c}\includegraphics[scale=0.45]{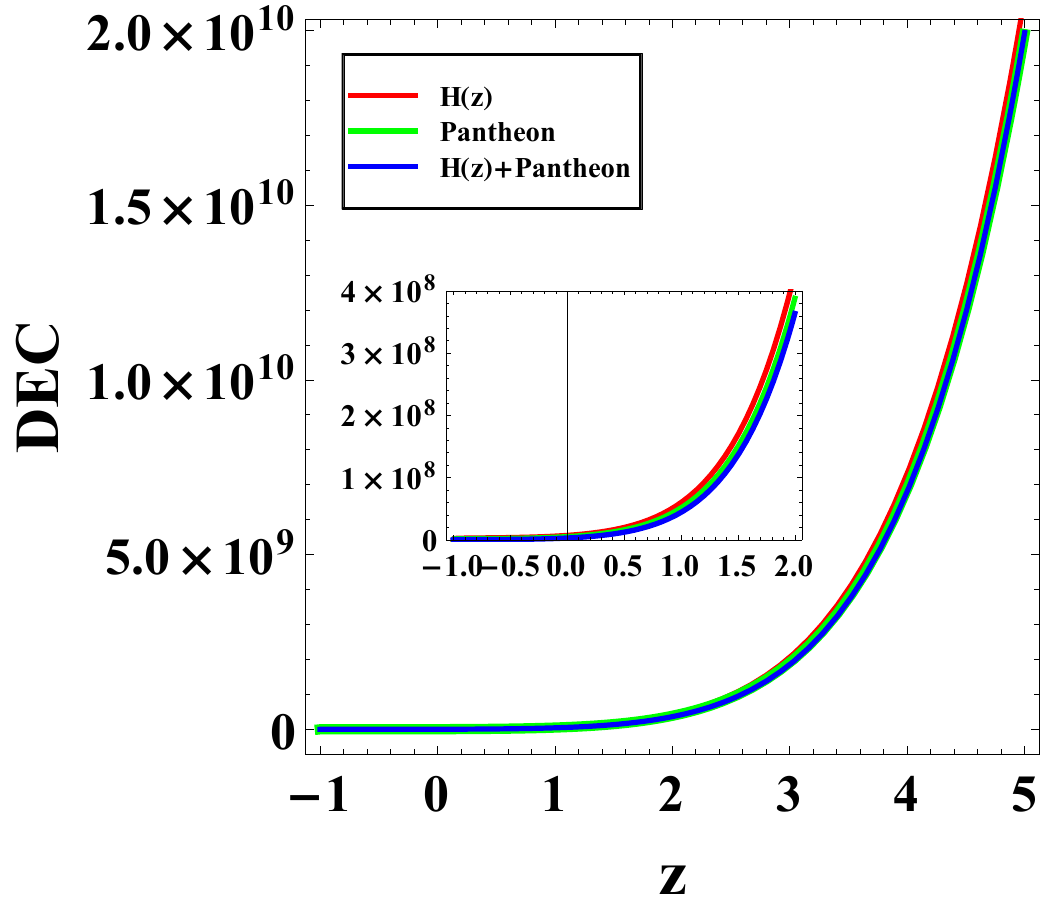}}
	\caption{The representative graphical plots of the energy conditions.}
        \label{ECs}
\end{figure}

In Fig. \ref{ECs}, we interestingly perceive that for all the observational datasets, NEC and SEC behave alike. Both the energy conditions hold good in the early universe but are violated in the present as well as in later eras. On the other hand, DEC holds for all time periods. The violation of SEC hints at the existence of dark energy in our model (see Fig. \ref{ECs}(b)). The breach of NEC indicates the phantom behavior of the model in late times, indicating its instability for an interval of redshift values (see Fig. \ref{ECs}(a)).

\section{Conclusion}\label{results}
\qquad In the present investigation, we try to investigate a cosmological model in the alternative gravity theory $f(R,T)$. The motivation for choosing this gravity theory arises from its capacity to explore both the physical and geometric behavior of the cosmos. We obtained the EFEs by taking a specific form of the $f(R,T)$ function. The Hubble parameter has been calculated here in terms of the redshift parameter along with three model parameters from the jerk and deceleration parameter relation. Also, we enumerate the best-fit values of the model parameters, namely, $j$, $H_0$, and $\alpha$, by utilizing the observational datasets $H(z)$, $Pantheon$, and their joint data (refer to Table \ref{tabparm}). For this purpose, we used the MCMC method with the help of the $emcee$ package in Python. Interestingly, one can find that our results of $ j $ and $ H_0 $ are consistent with the recently obtained Planck data. The contours plotted with $1-\sigma$ and $2-\sigma$ confidence levels have been discussed (see Figs. \ref{hzc}, \ref{panc}, and \ref{hzpanc}). To assess the performance of our $f(R,T)$ model against the $ \Lambda $CDM model, we have used error bar plots in Fig. \ref{err}. We observe the cosmic phase transition from deceleration to acceleration mode, marked by the shift of the $q$ from positive to negative values.
           
The top left panel of Fig. \ref{phypara} provides a visual narrative of the evolution of the energy density from the earlier to the later phase of the universe. For all three datasets, the curves of energy density decrease monotonically from high to low redshift. In Fig. \ref{phypara}(b), we notice that the isotropic pressure shows positive behavior in the early cosmic epoch, whereas it has a negative signature in the present stage and zero in the late epoch. This indicates an accelerating as well as expanding universe at present and a pressureless universe in the future. The trajectories of EoS parameter $ \omega $ pass through the various stages of evolution of the cosmos as it starts as perfect fluid enters the quintessence dark energy region and then after remains in the phantom region (see Fig. \ref{phypara}(c)).

In this model, we observe that the energy conditions, NEC and SEC, face violations at low redshift values, whereas DEC holds across the entire value of redshift (see Fig. \ref{ECs}).
    
Finally, based on the above-mentioned features, one can note that the presented model of $f(R,T)$ gravity in the background of the flat FLRW metric is quite different from the usual dark energy models. Undoubtedly, via the plots and exhibited physical features, our obtained model in the $f(R,T)$ gravity theory emerges as a promising candidate among various modified gravity theories and thus indicates the usefulness of further scope to work in the modified theory of gravity for verification of observational signatures in a more detailed and significant way.

\vskip0.2in

\section*{Acknowledgement} SR gratefully acknowledges support from the Inter-University Centre for Astronomy and Astrophysics (IUCAA), Pune, India, under its Visiting Research Associateship Programme as well as the facilities under ICARD, Pune, at CCASS, GLA University, Mathura, India. 

\section*{Data Availability Statement} In this manuscript, we have used observational data as available in the literature; as such, our work does not produce any form of new data.

\section*{Conflicts of Interest} The authors assert that there are no conflicts of interest of the publication of this work.


\begin{thebibliography}{99}

\bibitem{Capozziello:2002rd}
S.~Capozziello,
Int. J. Mod. Phys. D \textbf{11}, 483-492 (2002).

\bibitem{Nojiri:2003ft}
S.~Nojiri and S.~D.~Odintsov,
Phys. Rev. D \textbf{68}, 123512 (2003).

\bibitem{Carroll:2003wy}
S.~M.~Carroll, V.~Duvvuri, M.~Trodden and M.~S.~Turner,
Phys. Rev. D \textbf{70}, 043528 (2004).

\bibitem{Bertolami:2007gv}
O.~Bertolami, C.~G.~Boehmer, T.~Harko and F.~S.~N.~Lobo,
Phys. Rev. D \textbf{75}, 104016 (2007).

\bibitem{Harko:2008qz}
T.~Harko,
Phys. Lett. B \textbf{669}, 376-379 (2008).

\bibitem{Harko:2014pqa}
T.~Harko,
Phys. Rev. D \textbf{90}, 044067 (2014).
\bibitem{ShaHar1}
J.~K.~Singh, H.~Balhara, Shaily, P.~Singh,
Astronomy and Computing \textbf{46}, 100795 (2024).

\bibitem{Bengochea:2008gz}
G.~R.~Bengochea and R.~Ferraro,
Phys. Rev. D \textbf{79}, 124019 (2009).

\bibitem{Linder:2010py}
E.~V.~Linder,
Phys. Rev. D \textbf{81}, 127301 (2010)
[erratum: Phys. Rev. D \textbf{82}, 109902 (2010)].

\bibitem{Boehmer:2011gw}
C.~G.~Boehmer, A.~Mussa and N.~Tamanini,
Class. Quant. Grav. \textbf{28}, 245020 (2011).

\bibitem{Bamba:2010wfw}
K.~Bamba, S.~D.~Odintsov, L.~Sebastiani and S.~Zerbini,
Eur. Phys. J. C \textbf{67}, 295-310 (2010).

\bibitem{Houndjo:2013us}
M.~J.~S.~Houndjo, M.~E.~Rodrigues, D.~Momeni and R.~Myrzakulov,
Can. J. Phys. \textbf{92}, 1528-1540 (2014).

\bibitem{Nojiri:2005jg}
S.~Nojiri and S.~D.~Odintsov,
Phys. Lett. B \textbf{631}, 1-6 (2005).

\bibitem{ShaHar}
J.~K.~Singh, Shaily, H.~Balhara, K.~Bamba, and J.~Jena,
Astronomy and Computing, https://doi.org/10.1016/j.ascom.2024.100790 (2024).

\bibitem{Sharif:2016xjv}
M.~Sharif and A.~Ikram,
Eur. Phys. J. C \textbf{76}, 640 (2016).

\bibitem{Shamir:2018qhq}
M.~F.~Shamir and M.~Ahmad,
Phys. Rev. D \textbf{97}, 104031 (2018).

\bibitem{Myrzakulov:2015qaa}
R.~Myrzakulov, L.~Sebastiani and S.~Vagnozzi,
Eur. Phys. J. C \textbf{75}, 444 (2015).

\bibitem{Li:2010cg}
B.~Li, T.~P.~Sotiriou and J.~D.~Barrow,
Phys. Rev. D \textbf{83}, 064035 (2011).

\bibitem{Capozziello:2011nr}
S.~Capozziello, M.~De Laurentis, S.~D.~Odintsov and A.~Stabile,
Phys. Rev. D \textbf{83}, 064004 (2011).

\bibitem{Das:2017rhi}
A.~Das, S.~Ghosh, B.~K.~Guha, S.~Das, F.~Rahaman and S.~Ray,
Phys. Rev. D \textbf{95}, 124011 (2017).

\bibitem{Deb:2017rhc}
D.~Deb, F.~Rahaman, S.~Ray and B.~K.~Guha,
Phys. Rev. D \textbf{97}, 084026 (2018).

\bibitem{Deb:2018sgt}
D.~Deb, S.~V.~Ketov, S.~K.~Maurya, M.~Khlopov, P.~H.~R.~S.~Moraes and S.~Ray,
Mon. Not. Roy. Astron. Soc. \textbf{485}, 5652-5665 (2019).

\bibitem{Deb:2018gzt}
D.~Deb, S.~V.~Ketov, M.~Khlopov and S.~Ray,
JCAP \textbf{10}, 070 (2019).

\bibitem{Deb:2017rhd}
D.~Deb, F.~Rahaman, S.~Ray and B.~K.~Guha,
JCAP \textbf{03}, 044 (2018).

\bibitem{Biswas:2018inc}
S.~Biswas, S.~Ghosh, S.~Ray, F.~Rahaman and B.~K.~Guha,
Annals Phys. \textbf{401}, 1-20 (2019).

\bibitem{Biswas:2020gzd}
S.~Biswas, D.~Shee, B.~K.~Guha and S.~Ray,
Eur. Phys. J. C \textbf{80}, 175 (2020).

\bibitem{Biswas:2021wfn}
S.~Biswas, D.~Deb, S.~Ray and B.~K.~Guha,
Annals Phys. \textbf{428}, 168429 (2021).

\bibitem{Tripathy:2021vjt}
S.~K.~Tripathy, B.~Mishra, M.~Khlopov and S.~Ray,
Int. J. Mod. Phys. D \textbf{30}, 2140005 (2021).

\bibitem{Maurya:2021aio}
S.~K.~Maurya, F.~Tello-Ortiz and S.~Ray,
Phys. Dark Univ. \textbf{31}, 100753 (2021).

\bibitem{Mishra:2020jjk}
B.~Mishra, F.~M.~Esmeili and S.~Ray,
Indian J. Phys. \textbf{95}, 2245-2254 (2021).

\bibitem{Malik:2021dub}
A.~Malik,
Eur. Phys. J. Plus \textbf{136}, 1146 (2021).

\bibitem{Singh:2022eun}
J.~K.~Singh, A.~Singh, G.~K.~Goswami and J.~Jena,
Annals Phys. \textbf{443}, 168958 (2022).

\bibitem{Buchdahl:1970ldb}
H.~A.~Buchdahl,
Mon. Not. Roy. Astron. Soc. \textbf{150}, 1-8 (1970).

\bibitem{Nojiri:2007as}
S.~Nojiri and S.~D.~Odintsov,
Phys. Lett. B \textbf{657}, 238--245 (2007).

\bibitem{Nojiri:2007cq}
S.~Nojiri and S.~D.~Odintsov,
Phys. Rev. D \textbf{77}, 026007 (2008).

\bibitem{Cognola:2007zu}
G.~Cognola, E.~Elizalde, S.~Nojiri, S.~D.~Odintsov, L.~Sebastiani and S.~Zerbini,
Phys. Rev. D \textbf{77}, 046009 (2008).

\bibitem{Elizalde:2010ts}
E.~Elizalde, S.~Nojiri, S.~D.~Odintsov, L.~Sebastiani and S.~Zerbini,
Phys. Rev. D \textbf{83}, 086006 (2011).

\bibitem{Nojiri:2010wj}
S.~Nojiri and S.~D.~Odintsov,
Phys. Rept. \textbf{505}, 59--144 (2011).

\bibitem{Chakraborty:2014xla}
S.~Chakraborty and S.~SenGupta,
Eur. Phys. J. C \textbf{75}, 11 (2015).

\bibitem{Rodrigues:2012qua}
M.~E.~Rodrigues, M.~J.~S.~Houndjo, D.~Saez-Gomez and F.~Rahaman,
Phys. Rev. D \textbf{86}, 104059 (2012).

\bibitem{Salako:2013gka}
I.~G.~Salako, M.~E.~Rodrigues, A.~V.~Kpadonou, M.~J.~S.~Houndjo and J.~Tossa,
JCAP \textbf{11}, 060 (2013).

\bibitem{Paliathanasis:2016vsw}
A.~Paliathanasis, J.~D.~Barrow and P.~G.~L.~Leach,
Phys. Rev. D \textbf{94}, 023525 (2016).

\bibitem{Harko:2011kv}
T.~Harko, F.~S.~N.~Lobo, S.~Nojiri and S.~D.~Odintsov,
Phys. Rev. D \textbf{84}, 024020 (2011).

\bibitem{Houndjo:2011tu}
M.~J.~S.~Houndjo,
Int. J. Mod. Phys. D \textbf{21}, 1250003 (2012).

\bibitem{Yousaf:2016lls}
Z.~Yousaf, K.~Bamba and M.~Z.~u.~H.~Bhatti,
Phys. Rev. D \textbf{93}, 124048 (2016).

\bibitem{Shabani:2014xvi}
H.~Shabani and M.~Farhoudi,
Phys. Rev. D \textbf{90}, 044031 (2014).

\bibitem{Shabani:2013djy}
H.~Shabani and M.~Farhoudi,
Phys. Rev. D \textbf{88}, 044048 (2013).

\bibitem{Moraes:2015uxq}
P.~H.~R.~S.~Moraes, J.~Arba\~nil, D.V. and M.~Malheiro,
JCAP \textbf{06}, 005 (2016).

\bibitem{Roshan:2016mbt}
M.~Roshan and F.~Shojai,
Phys. Rev. D \textbf{94}, 044002 (2016).

\bibitem{Noureen:2015nja}
I.~Noureen and M.~Zubair,
Eur. Phys. J. C \textbf{75}, 62 (2015).

\bibitem{Moraes:2019hgx}
P.~H.~R.~S.~Moraes, P.~K.~Sahoo and S.~K.~J.~Pacif,
Gen. Rel. Grav. \textbf{52}, 32 (2020).

\bibitem{Shamir:2015rva}
M.~F.~Shamir,
Eur. Phys. J. C \textbf{75}, 354 (2015).

\bibitem{Singh:2014bha}
C.~P.~Singh and P.~Kumar,
Eur. Phys. J. C \textbf{74}, 3070 (2014).

\bibitem{Jamil:2011ptc}
M.~Jamil, D.~Momeni, M.~Raza and R.~Myrzakulov,
Eur. Phys. J. C \textbf{72}, 1999 (2012).

\bibitem{Correa:2015qma}
R.~A.~C.~Correa and P.~H.~R.~S.~Moraes,
Eur. Phys. J. C \textbf{76}, 100 (2016).

\bibitem{Sharif:2012zzd}
M.~Sharif and M.~Zubair,
JCAP \textbf{03}, 028 (2012)
[erratum: JCAP \textbf{05}, E01 (2012)].

\bibitem{Deng:2014hga}
X.~M.~Deng and Y.~Xie,
Int. J. Theor. Phys. \textbf{54}, 1739--1749 (2015).

\bibitem{Singh:2023gxd}
J.~K.~Singh, Shaily, A.~Singh, A.~Beesham and H.~Shabani,
Annals Phys. \textbf{455}, 169382 (2023).

\bibitem{Singh:2022jue}
J.~K.~Singh, H.~Balhara, K.~Bamba and J.~Jena,
JHEP \textbf{03}, 191 (2023) [erratum: JHEP \textbf{04}, 049 (2023)].

\bibitem{Singh:2018xjv}
J.~K.~Singh, K.~Bamba, R.~Nagpal and S.~K.~J.~Pacif,
Phys. Rev. D \textbf{97}, 123536 (2018).

\bibitem{Elizalde:2018arz}
E.~Elizalde and M.~Khurshudyan,
Phys. Rev. D \textbf{99}, 024051 (2019).

\bibitem{Mahapatra:2024kfq}
P.~Mahapatra and P.~K.~Das,
Preprint arXiv:2401.01321 [astro-ph.CO].

\bibitem{Starobinsky:1980te}
A.~A.~Starobinsky,
Phys. Lett. B \textbf{91}, 99--102 (1980).

\bibitem{Chimento:2007da}
L.~Chimento and M.~I.~Forte,
Phys. Lett. B \textbf{666}, 205--211 (2008).

\bibitem{Shaily:2022enj}
Shaily, M.~Zeyauddin and J.~K.~Singh,
Preprint arXiv:220.05076 [gr-qc].

\bibitem{Pan-STARRS1:2017jku}
D.~M.~Scolnic \textit{et al.} [Pan-STARRS1],
Astrophys. J. \textbf{859}, 101 (2018).

\bibitem{Riess:1998dv}
A.~G.~Riess, R.~P.~Kirshner, B.~P.~Schmidt, S.~Jha, P.~Challis, P.~M.~Garnavich, A.~A.~Esin, C.~Carpenter, R.~Grashius and R.~E.~Schild, \textit{et al.}
Astron. J. \textbf{117}, 707--724 (1999).

\bibitem{Jha:2005jg}
S.~Jha, R.~P.~Kirshner, P.~Challis, P.~M.~Garnavich, T.~Matheson, A.~M.~Soderberg, G.~J.~M.~Graves, M.~Hicken, J.~F.~Alves and H.~G.~Arce, \textit{et al.}
Astron. J. \textbf{131}, 527--554 (2006).

\bibitem{Hicken:2009df}
M.~Hicken, P.~Challis, S.~Jha, R.~P.~Kirsher, T.~Matheson, M.~Modjaz, A.~Rest and W.~M.~Wood-Vasey,
Astrophys. J. \textbf{700}, 331--357 (2009).

\bibitem{Contreras:2009nt}
C.~Contreras, M.~Hamuy, M.~M.~Phillips, G.~Folatelli, N.~B.~Suntzeff, S.~E.~Persson, M.~Stritzinger, L.~Boldt, S.~Gonzalez and W.~Krzeminski, \textit{et al.}
Astron. J. \textbf{139}, 519--539 (2010).

\bibitem{SDSS:2014irn}
M.~Sako \textit{et al.} [SDSS],
Publ. Astron. Soc. Pac. \textbf{130}, 064002 (2018).

\bibitem{Singh:2022nfm}
J.~K.~Singh, Shaily, S.~Ram, J.~R.~L.~Santos and J.~A.~S.~Fortunato,
Int. J. Mod. Phys. D \textbf{32}, 2350040 (2023).

\bibitem{Gelman:1992zz}
A.~Gelman and D.~B.~Rubin,
Statist. Sci. \textbf{7}, (1992) 457--472.

\bibitem{Raychaudhuri:1953yv}
A.~Raychaudhuri,
Phys. Rev. \textbf{98}, 1123-1126 (1955).

\bibitem{Ray1957}
A.~Raychaudhuri,
Z. Astrophys. \textbf{43}, 161 (1957).

\bibitem{Bhattacharyya:2021djv}
I.~Bhattacharyya and S.~Ray,
Int. J. Mod. Phys. D \textbf{30}, 2150092 (2021).

\bibitem{Das:2023umo}
S.~Das, A.~Panda, G.~Manna and S.~Ray,
Fortsch. Phys. \textbf{71}, 2200193 (2023).

\bibitem{Panda:2023jwe}
A.~Panda, S.~Das, G.~Manna and S.~Ray,
Phys. Dark Univ. \textbf{43}, 101397 (2024).


\end{thebibliography}
\end{document}